\documentclass[12pt]{iopart}

\usepackage{color}
\usepackage{graphicx}
\usepackage{float}
\usepackage{epsfig}
\usepackage{epstopdf}
\usepackage{subfig}
\usepackage{bm} 
\usepackage{hyperref}

\def\bB{{\bm B}}

\def\bb{{\bm b}}

\def\bk{{\bm k}}

\def\bkappa{{\bm \kappa}}

\def\bxi{{\bm \xi}}

\def\calO{{\cal O}}

\def\div{\nabla\cdot}
\def\grad{\nabla}

\def\del{\nabla}
\def\del2{\nabla^2}

\def\curl{\nabla\times}

\def\fsAver#1{\left\langle #1 \right\rangle} 

\def\begeqn{\begin{equation}}
\def\endeqn{\end{equation}}
\def\begeqnar{\begin{eqnarray}}
\def\endeqnar{\end{eqnarray}}
\def\begeqnarn{\begin{eqnarray*}}
\def\endeqnarn{\end{eqnarray*}}

\def\xiddot{{\ddot \xi}}
\def\bQ{\bm Q}

\begin{document}
\title{Dynamics of an $n=1$ explosive instability and its role in high-$\beta$ disruptions}
\author{A. Y. Aydemir,  B. H. Park and Y. K. In}
\address{National Fusion Research Institute, Daejeon 34133, Korea}                        
\eads{aydemir@nfri.re.kr}

\begin{abstract}
Some low-$n$ kink-ballooning modes not far from marginal stability are shown to exhibit a bifurcation between two very distinct nonlinear paths that depends sensitively on the background transport levels and linear perturbation amplitudes. The particular instability studied in this work is an $n=1$ mode dominated by an $m/n=2/1$ component. It is driven by  a large pressure gradient in weak magnetic shear and can appear in various high-$\beta,$ hybrid/advanced scenarios. Here it is investigated in reversed shear equilibria where the region around the safety-factor minimum provides favorable conditions.
For a certain range of parameters, a relatively benign path results in a saturated ``long-lived mode'' (LLM) that causes little confinement degradation. At the other extreme, the quadrupole geometry of the $2/1$ perturbed pressure field  evolves into a ballooning finger that subsequently transitions from exponential to explosive growth. The finger eventually  leads to a fast disruption with precursors too short for any mitigation effort.
Interestingly, the saturated LLM state is found to be metastable; it also can be driven explosively unstable by  finite-amplitude perturbations. Similarities to some high-$\beta$ disruptions in reversed-shear discharges are discussed.
\end{abstract}


\begin{center}
\end{center}
\maketitle

\section{Introduction}

In tokamaks, crossing certain operational boundaries in plasma density ($n_e$)\cite{greenwald2002}, current ($I_p$)\cite{shafranov1970, wesson1978}, or pressure ($\beta$)\cite{hender2007} can lead to a disruption, a sudden and uncontrolled loss of thermal and magnetic energy in the plasma. Among these, high-$\beta$ disruptions are particularly challenging, not only because of the high thermal energy content of the plasma, but also because of their extremely fast time-scales in some cases.

Most high-$\beta$ disruptions are mediated by a (neoclassical) tearing mode (NTM), typically with the mode numbers $m=2, n=1,$ that for various reasons lock to the wall, grow in size and eventually cause a loss of confinement\cite{wade2007, deVries2011}. Even when they do not lead to disruptions, NTM's tend to degrade confinement significantly so that their avoidance or stabilization in the ITER ELMy H-mode baseline scenario has been a high-priority research item (see for example \cite{gŸnter2003, laHaye2006}). When the plasma $\beta$ is pushed higher beyond the no-wall limit in ``hybrid/advanced tokamak'' regimes, more dangerous $n=1$ kink modes can become unstable. In the presence of a close-fitting wall, these are generally transformed into slow-growing resistive wall modes (RWM's)\cite{freidberg2014}. Again, if they are allowed to lock to the wall, RWM's can lead to disruptions. Fortunately, plasma rotation\cite{bondeson1994, fitzpatrick1996, garofalo2002}, kinetic effects\cite{hu2004, liu2008, zheng2010}, coupled with feedback-control methods\cite{wade2007, chu2003, pustovitov2006}, can stabilize RWM's well above the no-wall $\beta$ limit.

Since both NTM's and RWM's grow on a slow, resistive time scale, disruptions caused by these modes are easily identified by their long precursors on various diagnostics. In fact, because of their relatively slow time scale,  these are precisely the type of disruptions that are targeted by various disruption mitigation schemes, which require at least a few 10's of milliseconds of warning time\cite{granetz2007, hollmann2015}.
As stated earlier, however, tokamaks disrupt for a wide variety of reasons (see for example \cite{deVries2011}), and  not all disruptions follow this slow path where their arrival is well-advertised in advance; some in fact occur with little warning. 
Unfortunately, their very fast time scale apparently makes detailed studies difficult, and it is likely that their rare appearance in the literature does not accurately reflect their actual frequency in the experiments.

There do exist some documented high-$\beta$ disruptions with precursors of the order of a millisecond or less. For example:  $\beta$-limit disruptions in TFTR due to toroidally localized ballooning modes in the presence of $n=1$ magnetohydrodynamic (MHD) activity \cite{fredrickson1996}, localized resistive interchange modes that couple to a global $n=1$ mode and lead to a disruption in negative central shear (NCS) discharges in DIII-D\cite{chu1996, jayakumar2002}, and disruptions following an internal transport barrier (ITB) collapse in JET\cite{paley2005}. In these discharges, some of the important details were clearly different: in TFTR, at least initially, the $q=1$ surface was involved, whereas DIII-D and JET presumably both had $q_{min} \simeq 2.$ But generally,  a large pressure gradient in regions of weak magnetic shear is believed to have played an essential role. Thus, one of the things we will do in this work will be a short review of the resistive and ideal stability of such configurations.

However, linear stability analysis alone cannot explain the fast time scale of these disruptions.
The mode that is involved has to be growing near Alfv\'enic rates to account for the time scale, but it is not clear how a discharge evolving on the slow transport time scale can generate an unstable mode with a near-Alfv\'enic growth rate without producing a long series of precursor oscillations during its sub-Alfv\'enic period.
 
The point raised above is in fact part of a more generic problem: If {\em an event} (e.g., a sawtooth crash, edge-localized mode (ELM) crash, disruption, etc.) that makes macroscopic changes in the state of a discharge in a time scale $\tau_e$ is attributed to some global instability, the instability growth rate $\gamma_e$ at the time the event is observed has to be commensurate with that time scale, i.e., we need to have $\gamma_e\sim\calO(1/\tau_e).$ We can safely assume $1/\gamma_e\sim\tau_e \ll \tau_t,$ where $\tau_t$ is a characteristic transport time scale; otherwise the ``event'' cannot be distinguished from ordinary transport. If we assume that the changes in the mode growth rate occur entirely because of modifications to the background equilibrium by transport, then there will necessarily be a long period during which $1/\tau_t < \gamma(t) < \gamma_e$, i.e., a time when the mode is growing faster than the transport rate but does not yet have the eventual ``crash'' rate.  But then we are faced with two related questions: (i) During this period, can the mode grow without being detected? The short answer is probably ``no,'' since it is hard to imagine how such a mode could avoid generating a long series of precursors during the period mentioned. (ii) Since it is now growing faster than the transport time scale, would it not ``self-stabilize'' and saturate without causing the ``event''  by modifying the equilibrium faster than transport processes? Here a general answer is again difficult, but a mode that depends very sensitively on local conditions for stability can probably ``self-stabilize'' and saturate more easily than a global mode like an $m=1$ resistive kink or one that is responsible for a major disruption. 

Thus, disruptions or other events that occur without long precursors seem to require a different evolution scenario than the one proposed above. Instead of the mode growth rate $\gamma(t)$ slowly evolving with the background equilibrium, we have to consider mechanisms that can make changes in $\gamma(t)$ at a rate much faster than expected from transport alone. 

A mechanism proposed by Hastie\cite{hastie1987} for fast sawtooth crashes and by Callen\cite{callen1999} for the DIII-D disruption mentioned earlier assumes that, in response to a linearly increasing plasma pressure driven by auxiliary heating, the growth rate of a pressure-driven mode grows as $\gamma(t) \propto \beta^{1/2} \propto t^{1/2},$ where the plasma $\beta$ is defined as $\beta=2\mu_0 \fsAver{p}/B^2$ and $\fsAver{.}$ denotes a volume average. Then for large times, the plasma displacement, assumed to satisfy $\xiddot = \gamma^2\xi$,  can be shown to evolve as $\xi(t) \propto \xi_0 \exp{(t/\tau)^{3/2}}$, where $\tau \simeq (\tau_{MHD}^2\tau_h)^{1/3}$. Here $\tau_{MHD}$ is the fast MHD time scale, $\tau_{MHD} = L/v_A,$ where $v_A = B/\sqrt{\mu_0\rho_m}$ is the Alfv\'en velocity, $L$ is a global length scale, and $\tau_h$ is the slow ``heating time,'' comparable to the ``transport time'' mentioned earlier,  $\tau_h \sim \tau_t$. Thus, this mechanism seems to lead to a faster-than-exponential growth and possibly explain the near-absence of precursors  before some sawtooth crashes or disruptions. However, Cowley\cite{cowley1997} has shown that, because of the large separation in the MHD and transport time scales, this path to super-exponential growth requires an unrealistically small initial perturbation and can be ruled out.

There are, however, nonlinear processes  in plasmas that can generate explosive (faster-than-exponential) growth while the underlying mode is still not far from marginal stability. In a numerical study of the semi-collisional/collisionless  $m=1$ mode using a reduced two-fluid model, nonlinearities involving the parallel pressure gradient were shown to give a near-exponential increase in the {\em growth rate} of the mode\cite{aydemir1992}, providing a possible explanation for precursor-less, fast sawtooth crashes. Similarly, Cowley and colleagues\cite{cowley1997, wilson2004, myers2013, cowley2015} have shown that the nonlinear evolution of pressure-driven modes can generate finite-time singularities, again demonstrating how a long period of precursors can be avoided during a fast disruptive event.

In this work we extend our study of a specific example\cite{aydemir2016}, a pressure-driven $n=1$ kink-ballooning mode that can continue to grow exponentially well into its nonlinear regime and become explosive with an apparent finite-time singularity at the end. We show that it can actually exhibit two very different types of nonlinear behavior depending on small differences in the assumed transport levels and linear perturbation amplitudes. In addition to the explosive behavior, it can also display a more benign evolution and saturate in a ``long-lived mode'' (LLM) with only minor confinement degradation\cite{leeSG2016}. The LLM itself is shown to be a meta-stable state; it can be pushed into the explosive regime with small changes in the transport coefficients or with a finite-size perturbation.

The experimental context for this computational study is KSTAR discharges with $q_{95}\simeq 7,~q_0\simeq 2$, and a low inductive current fraction, similar to some hybrid/advanced scenarios\cite{sips2005}. With on axis electron cyclotron resonance heating (ECRH), the pressure profile peaks and drives an ideal $m/n=2/1$ mode that saturates at a small amplitude. The resulting long-lived mode (LLM) survives many tens of seconds (as long as ECRH is maintained), with only a small effect on confinement\cite{leeSG2016}. Although there are no documented examples for the explosive version of this mode in KSTAR, in the absence of any detailed study of KSTAR disruptions, their existence cannot be ruled out. The computational tool used here is  the CTD code, which solves the nonlinear MHD equations in toroidal geometry (see \cite{aydemir2015} and the references therein).

Before moving on to a discussion of the nonlinear results, in the next section we briefly review the salient features of the linear stability of pressure-driven modes.

\section{Linear stability}
A general understanding of the stability of pressure-driven modes can be obtained from a cursory examination of the ideal MHD energy integral, written here in its ``intuitive form'' (plasma contribution only)\cite{freidberg2014, greene1968}:
\begeqnar
\delta W_p & = & \frac{1}{2}\int \left\{\frac{|\bQ_\perp|^2}{\mu_0} + \frac{B^2}{\mu_0}|\div\bxi_\perp + 2\bxi_\perp\cdot\bkappa|^2 + \gamma p |\div\bxi|^2 \right\}dV \nonumber \\
& + &  \frac{1}{2}\int\left\{ -2(\bxi_\perp\cdot\grad p)(\bxi_\perp\cdot\bkappa) - \frac{J_\parallel}{B}(\bxi_\perp\times\bB)\cdot\bQ_\perp \right\}dV,
\endeqnar
The largest stabilizing contribution tends to be the $|\bQ_\perp|^2$ term in the first integral representing the line-bending energy, where $\bQ=\curl\bxi\times\bB_0$ is the perturbed field.
The destabilizing pressure-gradient and parallel current terms are grouped together in the second integral. 
The pressure gradient makes a destabilizing contribution to $\delta W_p$ only in those regions where  the field line curvature $\bkappa\equiv \bb\cdot\grad\bb=(\mu_0/B^2)\grad_\perp(p+B^2/2\mu_0)$ is ``unfavorable,'' i.e., where $\bkappa\cdot\grad p > 0.$ By having the displacement vanish where the curvature is favorable, $\bkappa\cdot\grad p < 0$, the net destabilizing contribution from the pressure forces can be maximized. This is the path the ballooning modes take, but they pay a price in excess line-bending energy since the perturbation is not constant along the field lines.

Another path for pressure-driven instabilities opens up if the magnetic shear is weak in a region of finite width.
Simplifying and expanding $\bQ$ around a rational surface, we have  $\bQ_\perp \simeq \bxi_\perp(i\bk\cdot\bB_0)$, and
\begeqn
Q_\perp^2 \simeq \xi_\perp^2(m-nq)^2\simeq \xi_\perp^2(nq_s')^2(r-r_s)^2.
\endeqn
Thus, if the global shear $s\equiv rq'/q$ is weak enough in regions with strong pressure gradients, interchange-like modes become possible even in ``Mercier-stable'' equilibria with $q^2 > 1,$ first recognized by Zakharov\cite{zakharov1978}. In fact, a rational surface is not necessary for instability. With $q\simeq (m+\epsilon)/n,~0 < \epsilon\ll 1,$ and $s\ll 1,$ the line-bending energy can be minimized again since $Q_\perp^2 \simeq \xi_\perp^2\epsilon^2,$ which can be overcome by a strong-enough pressure gradient. 

This simple piece of physics, strong pressure drive coupled with weak shear, is behind the quasi-interchange mode\cite{wesson1986, aydemir1987a, hastie1988, waelbroeck1988} (for $q_{min}\simeq 1$) and the ``infernal'' modes\cite{manickam1987, charlton1989, charlton1990} (for $q_{min} > 1$), both pressure-driven modes in low-shear equilibria. The former was studied in the context of fast sawtooth crashes caused by an internal $m/n=1/1$ mode. The latter are particularly dangerous global modes that can be unstable much below the $n\rightarrow\infty$ ballooning limit and lead to major disruptions. They are typically thought of as ``low-$n$'' modes, but of course the same physics can also make the $n=1$ mode unstable, which will be the focus of this work.

For computational economy, our earlier work\cite{aydemir2016} focused on the nonlinear evolution of pressure-driven $n=1$ modes in circular geometry, in both monotonic and weakly-reversed $q$ profiles. Here we will extend it to non-circular geometry and provide a brief review of the linear properties of the relevant modes. Partly because of KSTAR's recent interest in  advanced scenarios with internal transport barriers (ITB's), we will mainly consider reversed-shear equilibria with $q_{min} > 2.$

\begin{figure}[htbp]
\begin{center}
\includegraphics[width=5.5in]{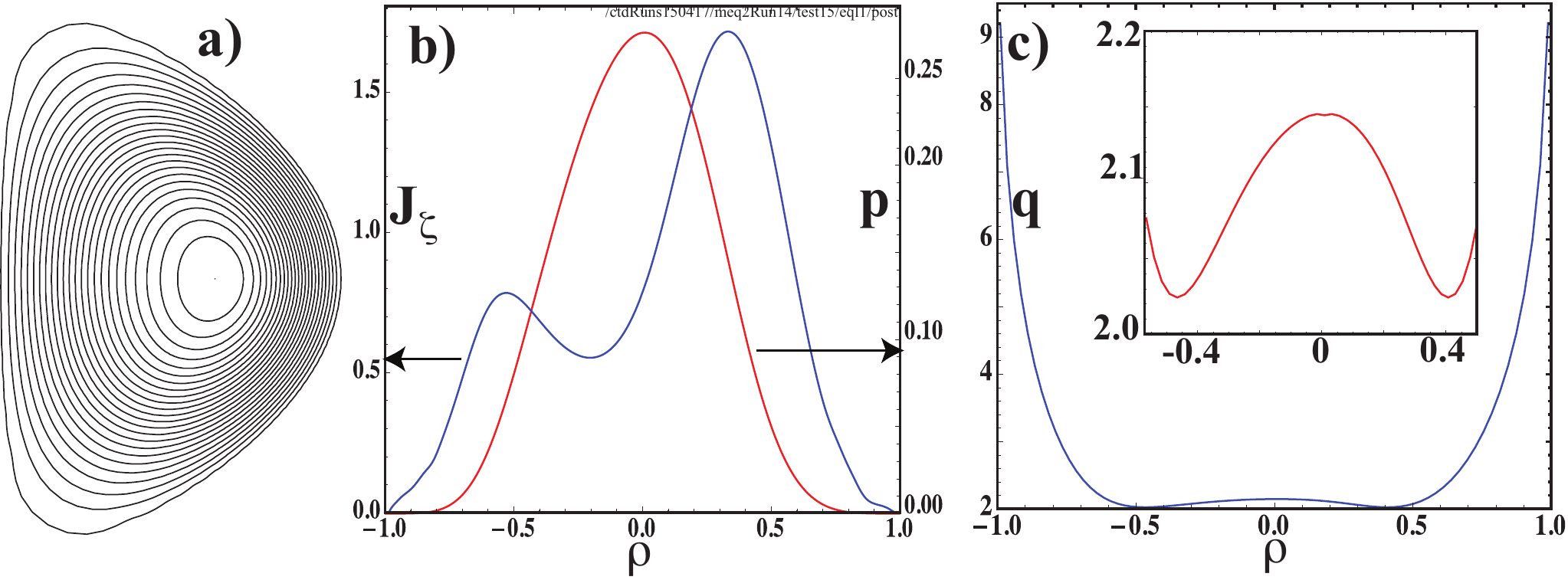}
\caption{\em \baselineskip 14pt Typical equilibrium profiles used in the linear and some of the nonlinear calculations. Here $q_0=2.15,~q_{min}=2.02,~q_l=9.21,~\beta_N=1.82,~\delta=0.6, \kappa=1.5$. (a) Flux surfaces. (b) Current density (left axis) and pressure (right axis) profiles. The horizontal coordinate $\rho$ is that of the $(\rho,\omega)$ conformal coordinates (see text). (c) Safety factor profile. The inset shows a magnified view of the central region. }
\label{fig:eql1}
\end{center}
\end{figure}

Typical equilibrium profiles used in the linear and some of the nonlinear calculations are shown in Fig.~\ref{fig:eql1}. The shaped geometry has $\kappa=1.5$ (elongation) and $\delta=0.6$ (triangularity) within a perfectly conducting boundary; these geometric parameters are held fixed, except when we revisit circular-geometry. The CTD code uses a conformal transform from the poloidal plane to a unit circle in $(\rho,\omega)$ coordinates to deal with weakly-shaped equilibria\cite{aydemir1990}. The coordinate axis is shifted to approximately align the $\rho=const.$ surfaces with flux surfaces, but $\rho$ is not a flux coordinate. For this reason, the plots as in Fig.~\ref{fig:eql1} (b,c) show both the $\omega=0$ (outboard) and $\omega=\pi$ (inboard) sections of the mid-plane. Note that a simple pressure profile without an internal or edge transport barrier is used to simplify the discussion.

As expected, resistivity enlarges the instability domain for the $n=1$ mode ($n>1$ stability is not considered in this work) so that an unstable mode is observed well below the ideal MHD stability limits. However, we find that the nature of the unstable resistive mode can be confusing. The ``infernal'' mode theory predicts a mode with a tearing scaling, $\gamma\tau_A \propto S^{-3/5},$ at low $\beta$. Close to the ideal stability boundary, the resistivity scaling is weaker, $\gamma\tau_A\propto S^{-3/13},$  becoming independent of $S$ beyond the ideal limit\cite{charlton1989}. 
Here we define $\tau_A$ as the shear-Alfv\'en time, $\tau_A = R_0/v_A,~v_A=B/\sqrt{\mu_0\rho_m}$. Then the magnetic Reynolds  (Lundquist) number is given by $S=\tau_R/\tau_A,$ where $\tau_R=\mu_0 a^2/\eta$ is the resistive diffusion time, and $a, R_0$ are the minor and major radii of the torus, respectively. 
Throughout this work, $S$ is defined in terms of the value of resistivity at the coordinate axis, but the resistivity itself is in general a function of the poloidal coordinates such that $\eta(\rho,\omega) J_{\zeta 0}(\rho,\omega)=E_0=const.$ The electric field $E_0$ is associated with the ``loop voltage.'' Numerically it is used to prevent the Ohmic diffusion of the equilibrium current during long nonlinear calculations.

The well-known interchange theory also predicts a resistive mode with the usual interchange scaling, $\gamma\tau_A \propto S^{-1/3}$\cite{glasser1975}, but only for reversed-shear equilibria. If we briefly recall the relevant theory, the Mercier (ideal interchange) modes are unstable for $D_I \equiv D_M - 1/4 > 0,$ where $D_M = -(2\mu_0rp'/B_\zeta^2s^2)(1-q^2)$ in circular geometry\cite{shafranov1968}. Although rare, Mercier modes have actually been observed experimentally\cite{in2000}.  Resistive instabilities require $D_R \equiv D_I + (H-1/2)^2 > 0,$ where $H\propto -p'/q'.$ We see that $D_R > 0$ is possible, even when Mercier stable ($D_I < 0$), with weakly-reversed shear ($H < 0$)  at high enough $\beta.$ At lower $\beta$ this mode also reverts to the tearing scaling with $S^{-3/5}$.

\begin{figure}[htbp]
\begin{center}
\includegraphics[width=6.0in]{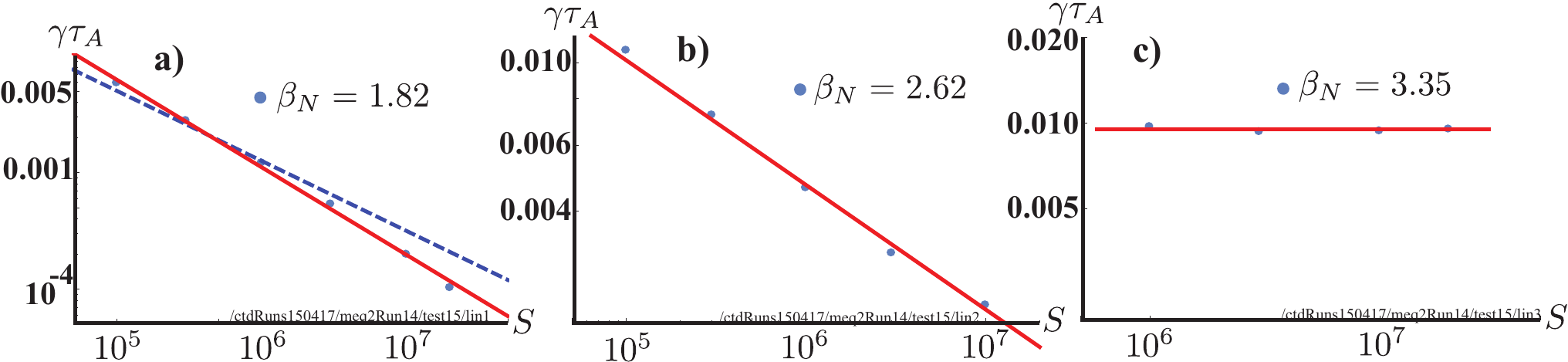}
\caption{\em \baselineskip 14pt Growth rate of the $n=1$ mode as a function of the magnetic Reynolds number $S$ for various values $\beta_N,$ for $q_0=2.15, q_{min}=2.02, q_l=9.21$ (see Fig.~\ref{fig:eql1}). (a) The blue dashed line represents the $S^{-3/5}$ tearing mode scaling, but $S^{-3/4}$ (red line) fits the computational data better (see text). (b) The data exhibits $S^{-1/3}$ scaling expected from resistive interchange modes. (c) At this $\beta_N$ the mode is ideally unstable. }
\label{fig:lin1-3}
\end{center}
\end{figure}

Stability of the equilibrium in Fig.~\ref{fig:eql1} is summarized in Fig.~\ref{fig:lin1-3}, where we plot the growth rate of the $n=1$ mode as a function of the magnetic Reynolds number $S$ for various values of the normalized $\beta,~\beta_N=\beta({\bar a}B)/I_p.$ Here ${\bar a} = a[(1+\kappa^2)/2]^{1/2}$ is an equivalent minor radius defined for an equilibrium with elongation $\kappa$, and $I_p$ is the plasma current. The $S$-scans are performed at a constant magnetic Prandtl number, $P_{M} = \mu/\eta=10,$ where $\mu$ is the normalized viscosity coefficient. Although normalized viscosity tends to be higher than resistivity in fusion plasmas, this value of $P_M$ is chosen entirely for numerical reasons.

At $\beta_N=1.82$ (Fig.~\ref{fig:lin1-3} (a)), there is a weakly unstable resistive mode; both the infernal mode and the resistive interchange theory seem to predict here a tearing-like scaling with $\gamma\tau_A \propto S^{-3/5}$ (the blue dashed line), but we find that a stronger dependence with $\gamma\tau_A \propto S^{-3/4}$ (the red line) is a better fit to the numerical data. Neither one of these theories takes into account viscous effects. The classical viscous-tearing mode theory that assumes $P_M < 1$ predicts a mode with the $S^{-2/3}$ scaling\cite{porcelli1987}, which is somewhat weaker than the $S^{-3/4}$ scaling we observe. It is possible that for $P_M \gg 1$, the $S^{-2/3}$ scaling changes to $S^{-3/4}$, but that possibility has not been investigated.

At $\beta_N=2.62$ (Fig.~\ref{fig:lin1-3} (b)), the mode is still resistive and has a clear resistive interchange scaling, $\gamma\tau_A\propto S^{-1/3}$. Here it is possible that there is a resistive infernal mode (with the $S^{-3/13}$ scaling\cite{charlton1989}) that is in competition with the interchange, but it is not observed numerically. The weak reversed shear (not considered in the infernal mode theory) may be making the resistive interchange the dominant mode in this particular parameter regime. 

At an even higher $\beta$ ($\beta_N=3.35$, panel (c)), the mode is ideally unstable (with wall).  During this study, $\beta_N$ was increased in large steps so that the locations of the transition points between various regimes are not known, a task left for a future work.

\begin{figure}[htbp]
\begin{center}
\includegraphics[width=4.0in]{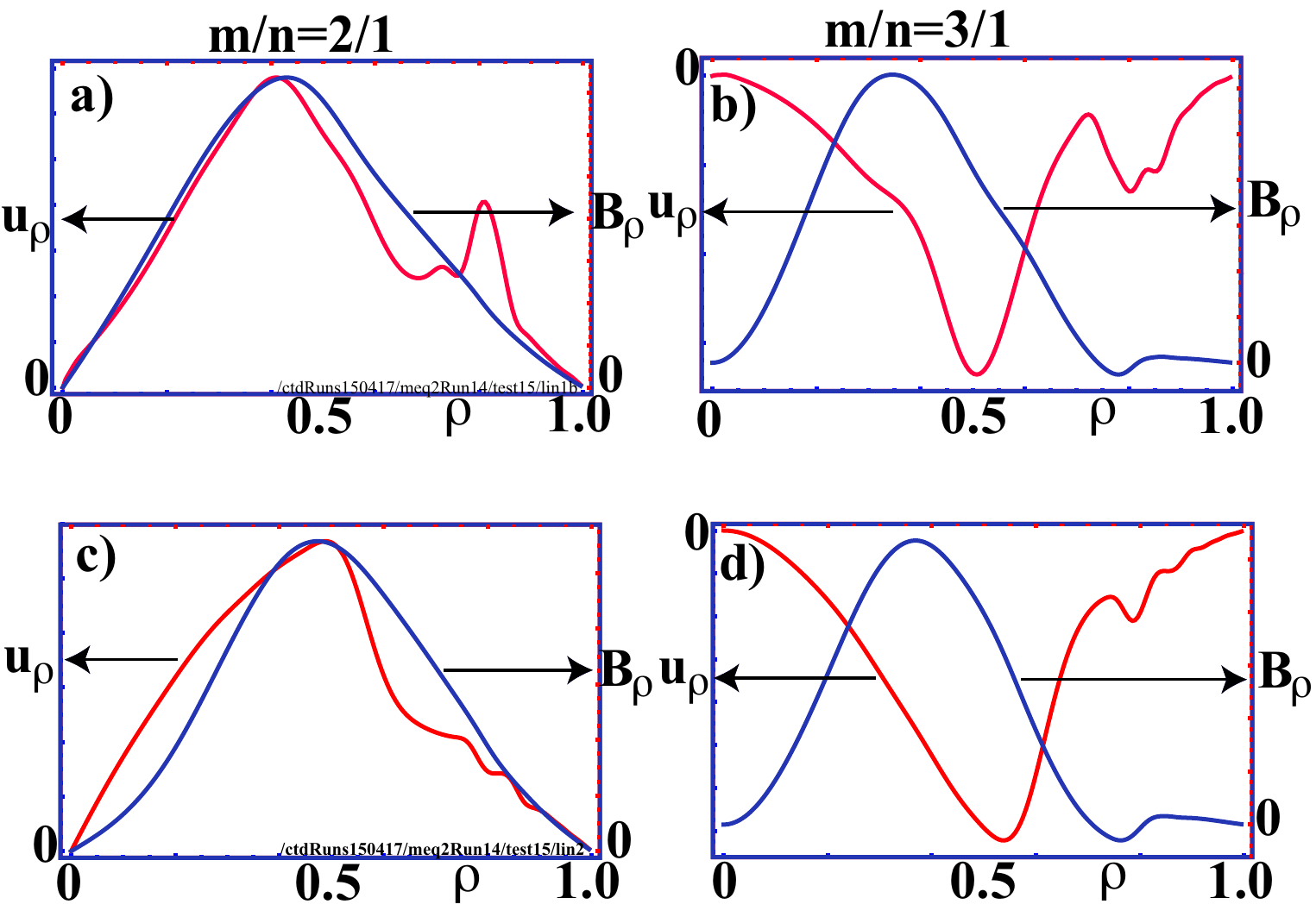}
\caption{\em \baselineskip 14pt Eigenfunctions for the radial velocity $u_\rho$ and magnetic field $B_\rho$, for $m/n=2/1$ and $m/n=3/1$ (in arbitrary units) at $S=10^6$ for the equilibrium shown of Fig.~\ref{fig:eql1}. (a,b) $\beta_N=1.82$, (c,d) $\beta_N=2.62.$ }
\label{fig:urBr1-2}
\end{center}
\end{figure}

Since there is no $q=2$ rational surface in the plasma for the series of equilibria considered here ($q_{min} >2$), and because of the wide region of weak magnetic shear around $q_{min}$ (see Fig.~\ref{fig:eql1}), the eigenfunctions do not exhibit a distinctive ``singular'' behavior there. In fact, they have the features of a global kink mode, as seen in Fig.~\ref{fig:urBr1-2}. For $\beta_N=1.82$ (panels (a,b)), there is a strong coupling to an $m/n=3/1$ mode near the edge, a $\Delta'$-stable tearing mode. At $\beta_N=2.62$ (panels (c,d)), this coupling seems to become weaker, probably explaining the change in scaling  from ``tearing''-like to $S^{-1/3}$ mentioned earlier. Note that although only two modes are shown, the linear calculations for $n=1$ included all poloidal mode numbers in the range $m \in [-7,57].$

Above we discussed in some detail the linear stability for $q_0=2.15,~q_{min}=2.02,$ mainly to place our nonlinear calculations below in some context. Summarizing our other linear results, for a more deeply-reversed equilibrium with $q_0=2.57,~q_{min}=2.02$ we find no ideal instability for  $\beta_N \le 3.08,$ the limit of our numerical explorations  for this $q$-profile. On the other hand, for an equilibrium with $q_0=2.03,~q_{min}=2.02$ (much weaker central shear), we find that $\beta_N \ge 2.72$ is ideally unstable, although the actual stability boundary has not been explored and is probably lower (but higher than $\beta_N=1.90$, where we find a resistive mode).

\section{The explosive instability and disruptions}
Nonlinearly the pressure-driven $n=1$ mode can turn into an explosive instability, as it was
first demonstrated in circular geometry\cite{aydemir2016}. Here we discuss the nonlinear evolution of the mode in shaped geometries using the linear results of the previous section as starting points. 

An ideally unstable  $n=1$ mode (e.g., one from Fig.~\ref{fig:lin1-3} (c)) with its large growth rate will naturally lead to a fast disruption. As discussed at some length in the Introduction, however, MHD modes are not ``born'' in this robustly unstable state. They tend to come into existence as weak resistive instabilities as the equilibrium slowly (on transport time scale) passes through some marginal stability point due to the evolving discharge conditions. Hence our goal in this section is to demonstrate how a weak resistive instability can evolve into a robust mode that will result in a fast disruption with only a brief period of precursors.

Thus we start with a weakly unstable equilibrium similar to that of Fig.~\ref{fig:eql1}; some of the relevant parameters are: $q_0=2.145, q_{min}=2.023, q_l=9.424, \beta_N=1.67$, which results in an even weaker instability than that of Fig.~\ref{fig:lin1-3}(a). The nonlinear calculations are performed at $S=10^6,~P_M=\mu/\eta=10,$ using 21 toroidal Fourier modes. The poloidal Fourier expansions have $m\in [0,64]$ for $n=0$ and $m\in [-5,64]$ for $n\in[1,20].$ The finite difference scheme in the radial direction uses $192$ grid points. Some of the algorithmic details of the CTD code used here can be found in \cite{aydemir2015} and the references therein.

\begin{figure}[htbp]
\begin{center}
\includegraphics[width=5.5in]{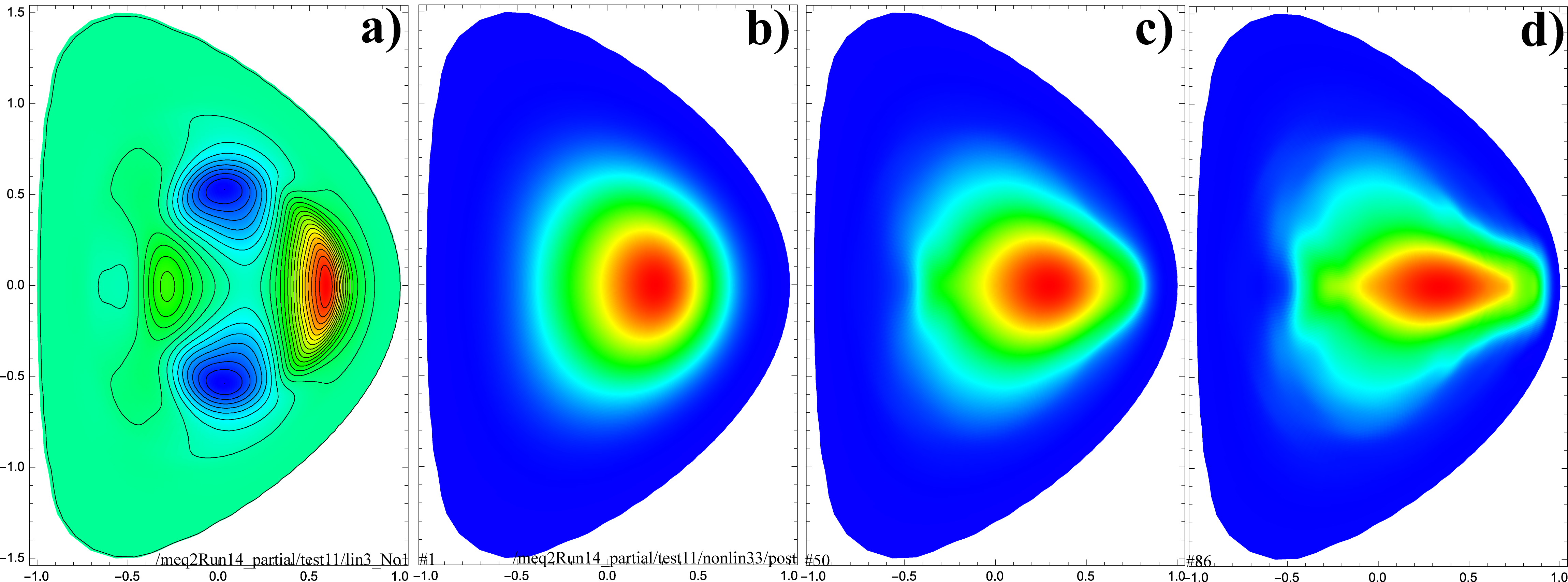}
\caption{\em \baselineskip 14pt The pressure field. (a) $n=1$ eigenfunction showing the dominance of the $m=2$ harmonic. The ballooning nature of the mode, with its in-out asymmetry, is easily visible. (b-d) Nonlinear development of the explosive finger. In units of the poloidal Alfv\'en time, $t_b=2.242\times 10^3, t_c=4.008\times 10^3, t_d=4.271\times 10^3.$ The figures are from the $\zeta=\pi$ plane (see also Fig.~\ref{fig:thetaZetaCombined} below).}
\label{fig:imagCombined}
\end{center}
\end{figure}

With weak shear and $q_{min} > 2$, an important  feature of the linear eigenfunction is the dominance of the $m=2$ poloidal component. This is clearly seen in Fig.~\ref{fig:imagCombined}(a), which shows the quadrupole geometry of the pressure perturbation (some coupling to an $m=3$ on the outside is also visible). 
This perturbation leads to an elliptical deformation of the flux surfaces in the core plasma that eventually forms a ballooning finger, as seen in Figs.~\ref{fig:imagCombined}(b-c). The finger pushes through the flux surfaces on near-Alv\'enic time scales and brings the core plasma in contact with the boundary (Fig.~\ref{fig:imagCombined}(d)). Using typical parameters for modern tokamaks ($B=3~T,~n_e=10^{19}~cm^{-3}, R_0=3~m$), the state with no visible deformation in Fig.~\ref{fig:imagCombined}(b)  and the final state in Fig.~\ref{fig:imagCombined}(d) are separated by less than $1~ms.$ Thus an actual disruption following this path would have a  very short warning time.

\begin{figure}[htbp]
\begin{center}
\includegraphics[width=5.5in]{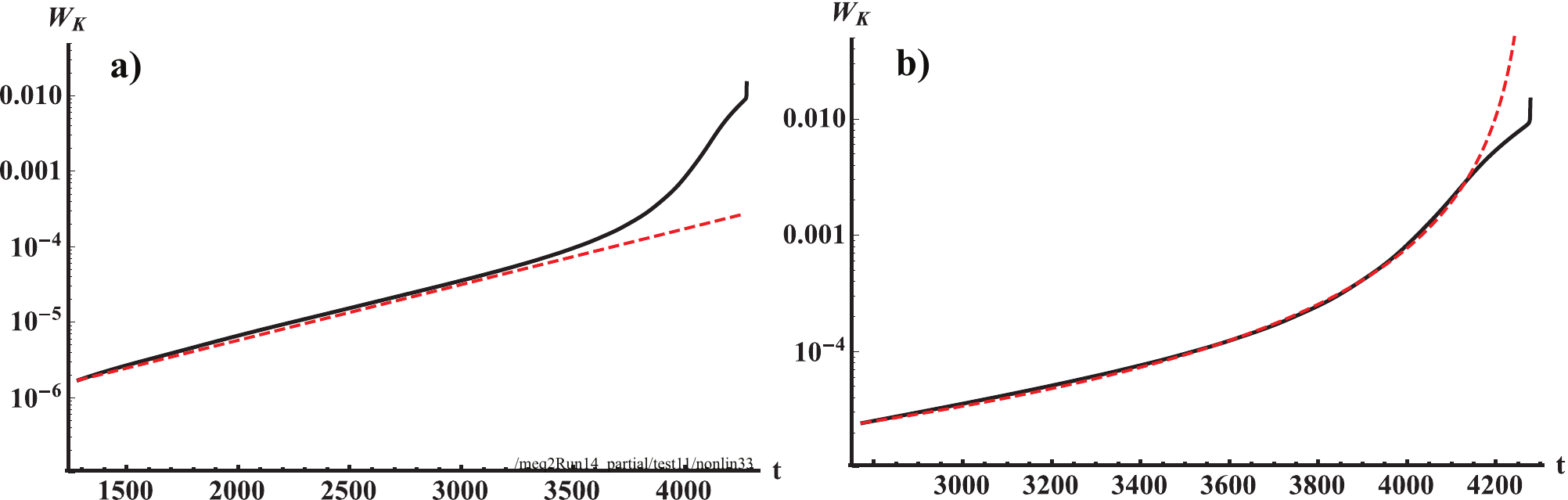}
\caption{\em \baselineskip 14pt Time history of the total kinetic energy in the $n\ge 1$ modes. (a) The dashed (red) line corresponds to a growth rate of $\gamma\tau_A=1.70\times 10^{-3}.$ Note that the growth becomes superexponential beyond $t\simeq 3000.$ (b) The finite-time singularity model of Eq.~\ref{eqn:finiteTime} is a good fit to the explosive phase (the dashed curve). The ballooning finger slows down beyond $t\simeq 4100$ as it starts coming into contact with the wall. }
\label{fig:wkCombined}
\end{center}
\end{figure}

The rapidity of the final disruptive phase is clearly due to a nonlinear increase in the growth rate rather than a slow, transport time scale change. This is seen in Fig.~\ref{fig:wkCombined} where the kinetic energy in the $n\ge 1$ modes (excluding the $n=0$ equilibrium flows) is plotted. Two points are immediately obvious: (i) The mode continues to exponentiate well into the nonlinear regime with a growth rate $\gamma\tau_A = 1.70\times 10^{-3}$ (the dashed red line in panel (a)). (ii) Instead of saturation, the late stages are characterized by a superexponential or explosive growth. In fact this phase has the appearance of a
a finite-time singularity (panel (b)) where the growth has the form
\begeqn
W_K(t)=W_K(t_i)[(t_f-t_i)/(t_f-t)]^\nu, \label{eqn:finiteTime}
\endeqn
where $t_i=2773.9,~t_f=4277.6,$ and the exponent $\nu=2.05.$ In our earlier calculations in circular geometry the explosive phase was even faster, with $\nu=3.37$\cite{aydemir2016}. The slowdown seen here has both physical and numerical sources: The underlying $n=1$ mode is inherently more stable in shaped geometry. And, because the poloidal spectrum for each toroidal mode is much wider in shaped geometry, fewer ($n_{max}=20$) toroidal modes were used here. In circular geometry we were able to use $n_{max}=30,$ which allowed us to continue the calculations further into the explosive phase.

\begin{figure}[htbp]
\begin{center}
\includegraphics[width=5in]{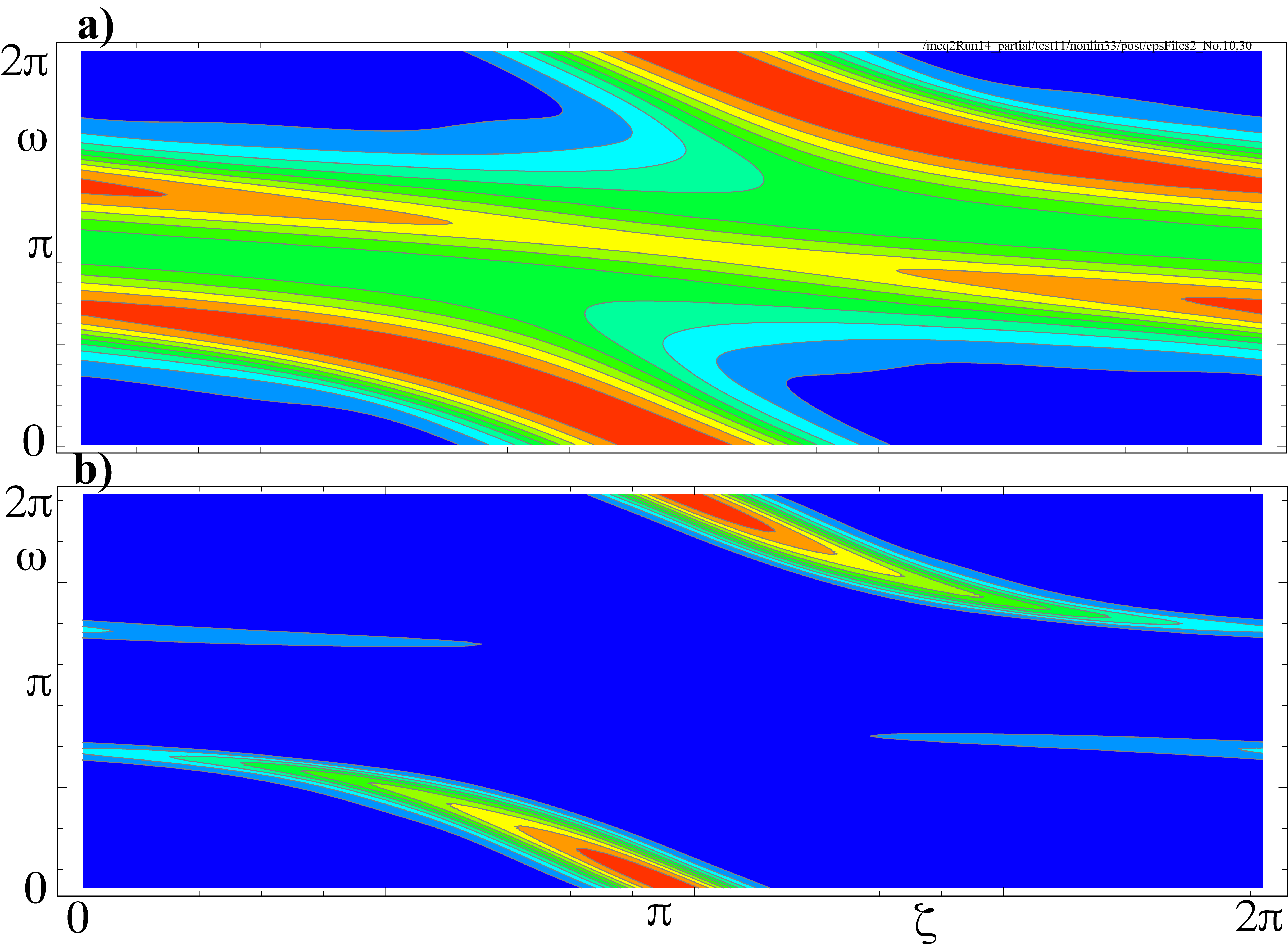}
\caption{\em \baselineskip 14pt Pressure contours in the $(\omega,\zeta)$ plane at $t=4270.9.$ The outboard (inboard) midplane is at $\omega=0~(\pi).$ Two  constant-$\rho$ surfaces are shown: (a) At $\rho=0.31$, the finger shows a ballooning structure but is almost completely extended around the torus along the field lines. (b) Further out at $\rho=0.84,$ the finger is more localized, in both parallel and perpendicular directions. Note that the $m/n=2/1$ helicity is preserved.}
\label{fig:thetaZetaCombined}
\end{center}
\end{figure}

The ballooning finger of Fig.~\ref{fig:imagCombined} has an extended structure along the field lines with $q_s=m/n=2/1$ helicity. As seen in Fig.~\ref{fig:thetaZetaCombined}, this symmetry is preserved as the finger moves outward. Because of magnetic shear, however,  it becomes more localized in both parallel (along the field) and perpendicular directions (compare panels (a) and (b)) as it moves through regions of increasing safety factor in order to reduce the line-bending energy. Thus, under experimental conditions when it eventually comes into contact with the wall or limiter, its thermal content will be deposited in a small area and possibly cause serious damage.

\begin{figure}[htbp]
\begin{center}
 { \subfloat{ \includegraphics[height=6.5cm]{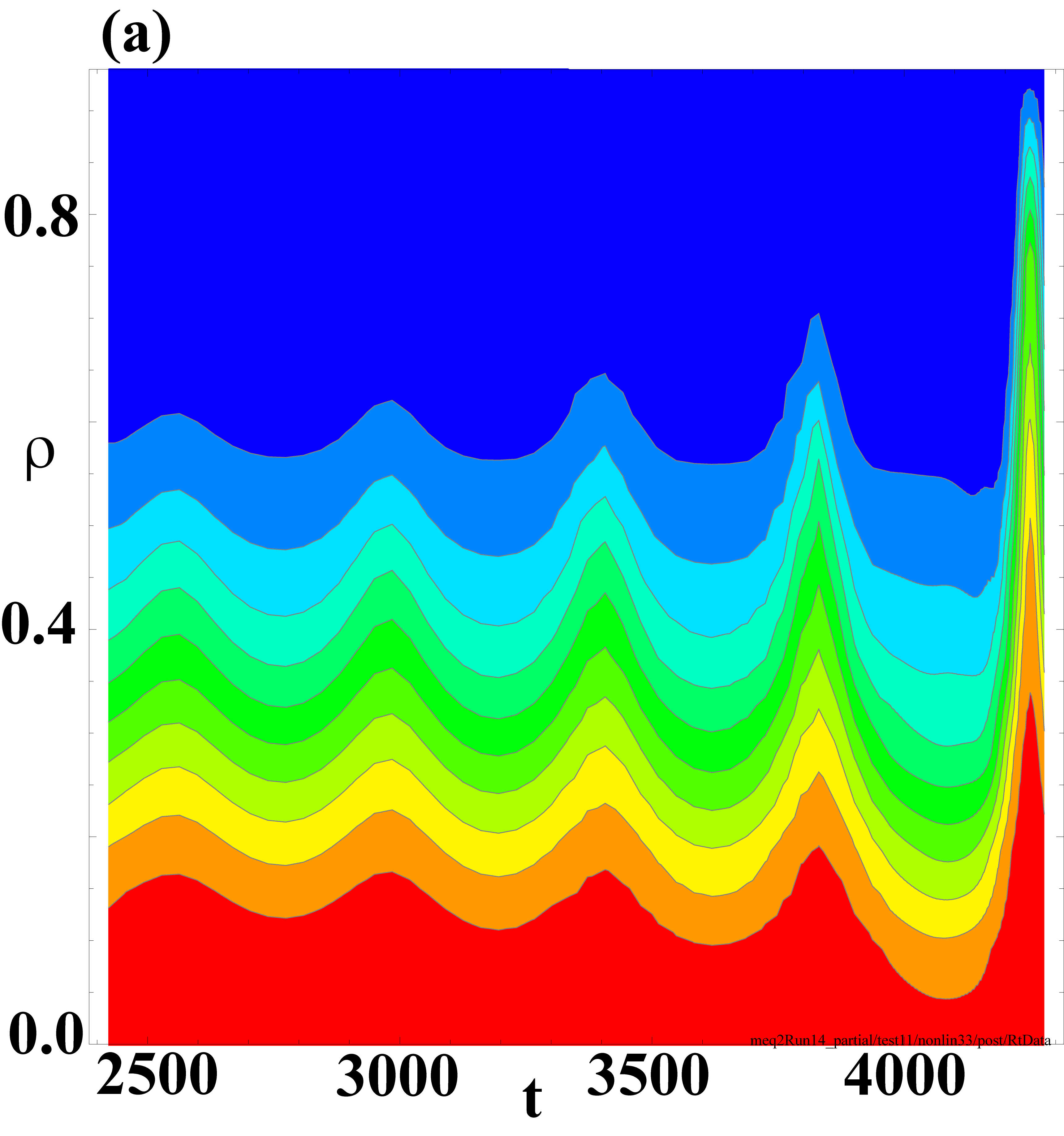}}
    \subfloat{ \includegraphics[height=6.5cm]{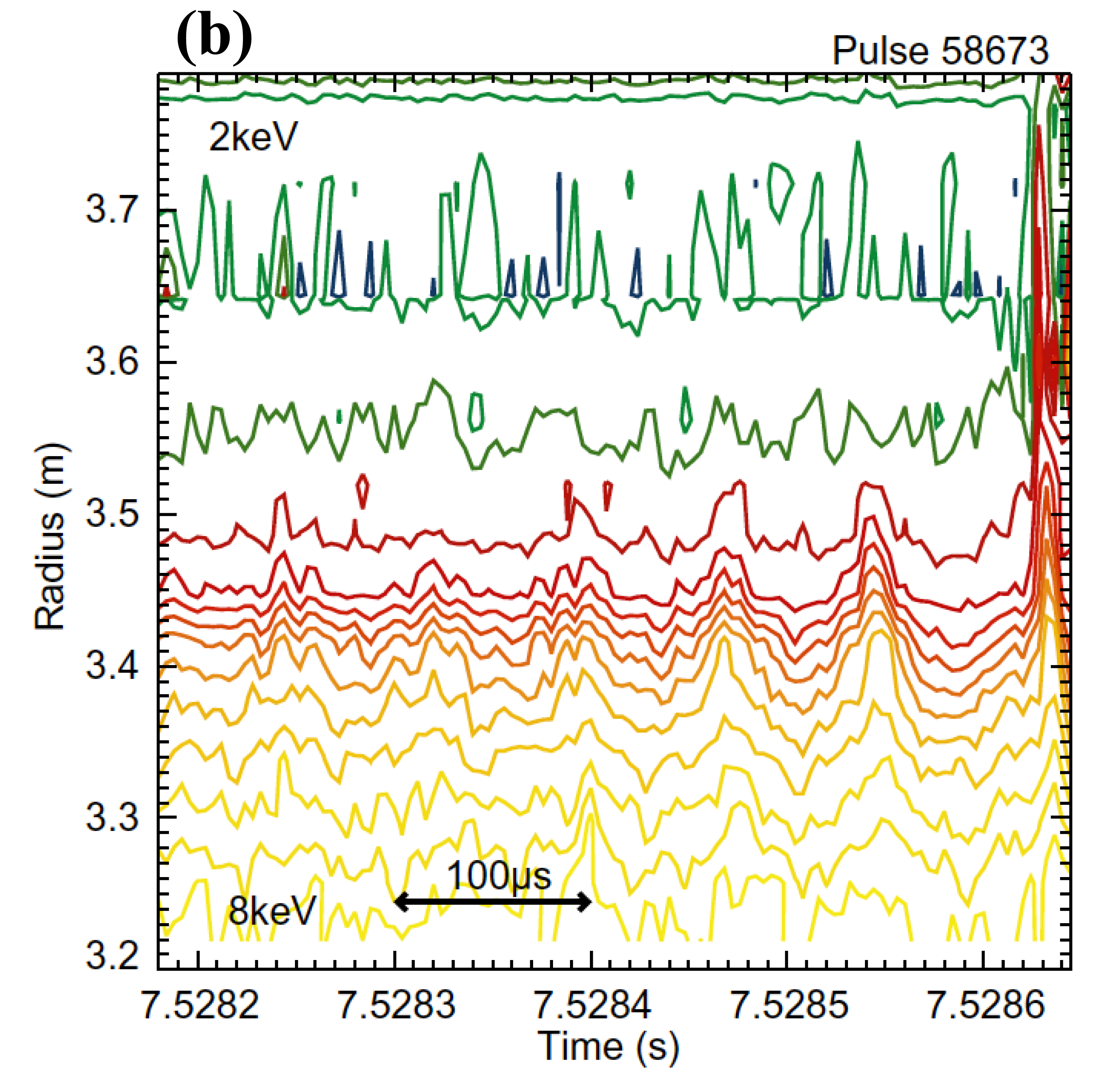}}
\caption{\em \baselineskip 14pt Comparison of synthetic ECE diagnostics with the JET data from Ref.~\cite{paley2005}. (a) Computational data with an assumed rigid toroidal rotation of period $425\tau_A.$ Note the narrowing of the finger in the time domain as it propagates radially outward, which of course corresponds to toroidal localization. (b) Experimental ECE contours from JET (Fig. 4 from Ref.~\cite{paley2005}, used with permission).}
\label{fig:RtPlot}    
 }
 \end{center}
\end{figure}

There have been observations of ballooning fingers during high-$\beta$ disruptions in  \break TFTR\cite{fredrickson1996, fredrickson1995}. JET also has reported similar results. During disruptions following an ITB collapse, localized disturbances in the ECE data that propagate from the ITB to the edge at velocities approaching 3km/s are seen\cite{paley2005}. These experimental observations can be  associated with the radial propagation of a ballooning finger as shown here.
In fact, a comparison of synthetic ECE diagnostics from our calculations with the JET data from Ref.~\cite{paley2005} (their Fig.~4) shows good agreement, as seen in Fig.~\ref{fig:RtPlot}.

Although there are several observations in KSTAR that may be associated with fast, high-$\beta$ disruptions without significant precursors, the necessary analysis to link them formally to an explosive instability has not been carried out. 

\section{Bifurcated states}
Generally, away from exact marginal points, we expect small changes in a relevant parameter to result in similarly small changes in the evolution of an unstable mode. Thus, small differences in the resistive dissipation level rarely have a significant impact on the saturation width of a tearing mode. However, there are counter examples where, for instance, an increase in the Prandtl number (ratio of viscous to resistive dissipation) beyond a threshold leads to a qualitatively different nonlinear regime\cite{maget2016}.

\begin{figure}[htbp]
\begin{center}
\includegraphics[width=5in]{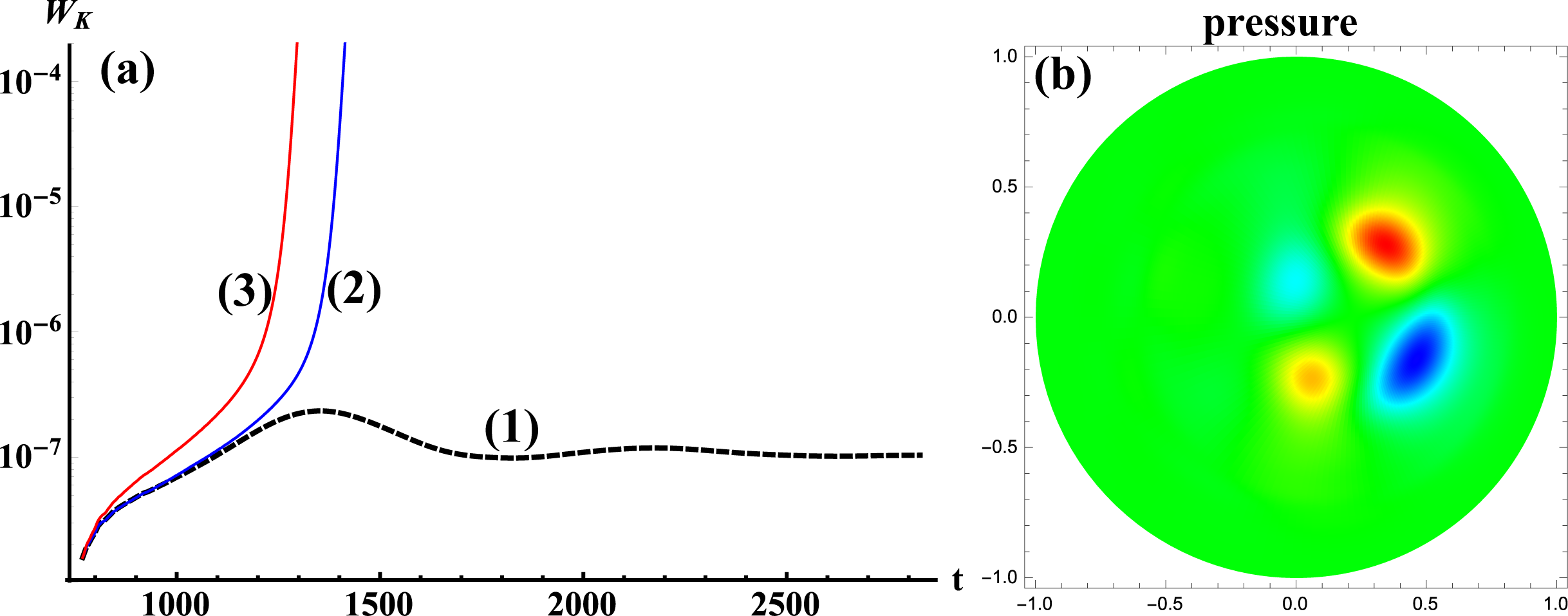}
\caption{\em \baselineskip 14pt (a) Time histories of the total kinetic energy in the $n\ge 1$ modes during the nonlinear evolution of the $n=1$ kink-ballooning mode. Curve (1): With thermal conductivity $\kappa_\perp=4.0\times 10^{-6}$ and viscosity $\mu=1\times 10^{-5}$ the mode saturates at a small amplitude resulting in a ``long-lived mode.'' Curve (2): $\kappa_\perp=3.5\times 10^{-6},~\mu=1\times 10^{-5}$ leads to explosive behavior where the energy displays a finite-time singularity. Curve (3): $\kappa_\perp=4\times 10^{-6},~\mu=6\times 10^{-6}$ is again explosive. (b) Saturated pressure field for the long-lived mode (curve  (1) in (a)). In order to bring out the details, the axisymmetric component ($n=0$) has been subtracted.}
\label{fig:bifurcation}
\end{center}
\end{figure}

Here we demonstrate an extreme case where a small change in a transport coefficient leads to a bifurcation between a benign, saturated state (the long-lived mode, LLM) and an explosive instability for the $n=1$ kink-ballooning mode. For computational economy, we expand upon our earlier results\cite{aydemir2016, leeSG2016} while staying in circular geometry.
The bifurcation is summarized in Fig.~\ref{fig:bifurcation} where we follow the nonlinear evolution of the mode starting with the same initial conditions and linear perturbation, but using slightly different transport coefficients. With $S=10^6$, thermal conductivity $\kappa_\perp=4\times 10^{-6}$ and viscosity $\mu=1\times 10^{-5}$, the mode goes through an exponential growth phase but saturates at a small amplitude (curve (1) in Fig.~\ref{fig:bifurcation}(a)). This regime is identified with the long-lived mode (LLM) observed in KSTAR, where an $m/n=2/1$ perturbation is seen in the electron cyclotron imaging (ECEI) data for many tens of seconds during the current flat-top period. The experimental  conditions under which the LLM was observed is described in more detail in \cite{leeSG2016, leeSG2016b}.

In a slightly less dissipative system with $\kappa_\perp=3.5\times 10^{-6}$ but $\mu=1\times 10^{-5}$ still, the mode initially follows a similar path and goes through an exponential-growth phase (curve (2) in (a)). However, instead of saturating, it gradually enters a super-exponential regime where the growth rate itself increases rapidly, eventually becoming explosive. A similar effect is seen with reduced viscosity: with $\mu=6\times 10^{-6},~\kappa_\perp=4\times 10^{-6}$  the mode again transitions into the explosive regime (curve (3)).

This bifurcation can be understood qualitatively if we assume the explosive phase has a finite threshold in the perturbation amplitude. Dissipation affects both the linear growth rate and the nonlinear saturation amplitude of the unstable mode. The higher dissipation level clearly causes the mode to saturate below the apparent threshold. This point is confirmed in the next section where we show that the curve (1) of Fig.~\ref{fig:bifurcation}(a) represents a continuous set of  metastable states.

\section{Metastability}
The  $n=1$ mode for the set of equilibria we consider here is {\em linearly} unstable, which implies that an infinitesimal perturbation will grow in time exponentially, at least until the mode attains a finite amplitude. The results of the previous section imply that whether it turns into a LLM or becomes explosive is determined by a critical perturbation amplitude that itself is a function of the dissipation coefficients, $\xi_{cr} = \xi_{cr}(\eta,\mu,\kappa_\perp)$. If it nonlinearly saturates below the threshold, $\xi < \xi_{cr}$, the result is a benign long-lived mode. Above $\xi_{cr}$ it becomes explosive. Thus, the pressure-driven $n=1$ mode  is said to exhibit {\em metastability}. In this section this theoretically predicted behavior\cite{cowley2015} is demonstrated numerically. 

\begin{figure}[htbp]
\begin{center}
\includegraphics[width=3.5in]{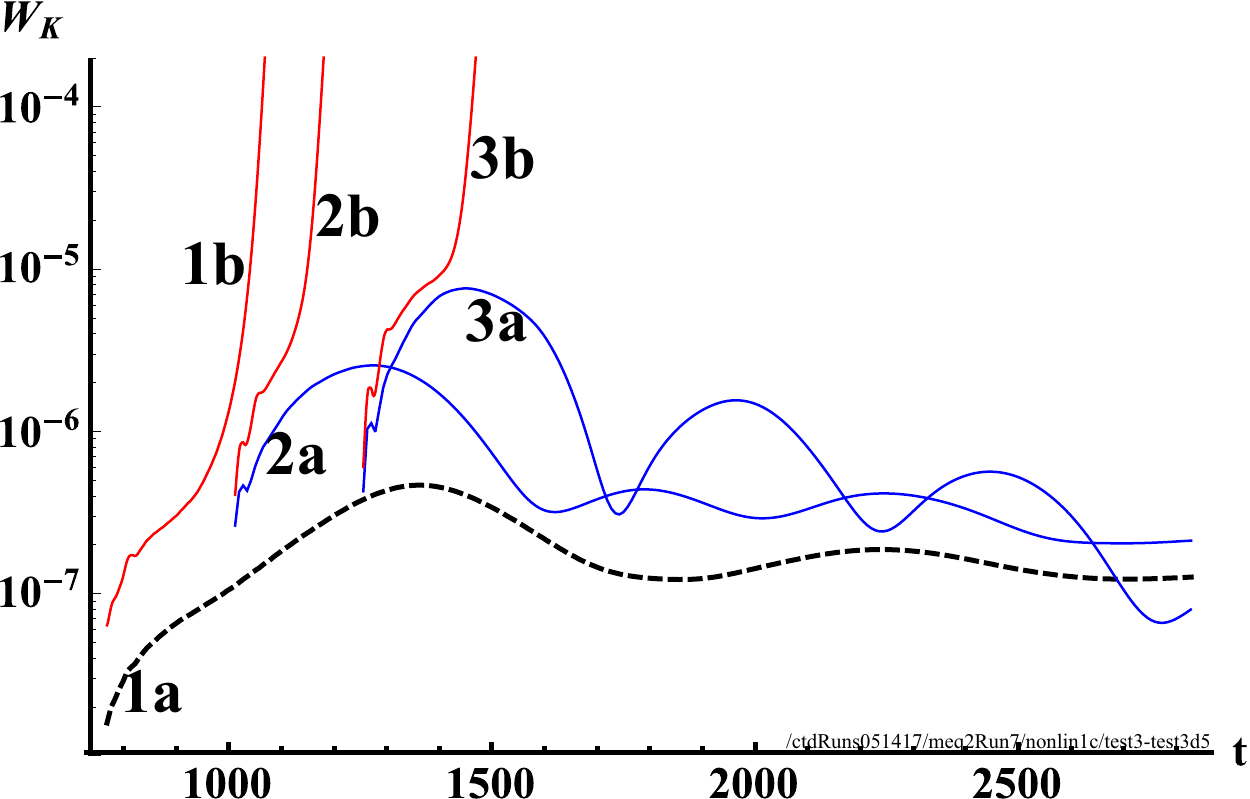}
\caption{\em \baselineskip 14pt Metastability of the long-lived mode (curve 1a) is demonstrated using perturbations of increasing amplitude. Curve 1a: $\epsilon=1\times 10^{-6}$ (stable). 1b: $\epsilon=2\times 10^{-6}$ (explosive). 2a: $\epsilon=4\times 10^{-6}$ (stable). 2b: $\epsilon=5\times 10^{-6}$ (explosive). 3a: $\epsilon=5\times 10^{-6}$ (stable). 3b: $\epsilon=6\times 10^{-6}$ (explosive). Here the parameter $\epsilon$ is a measure of the initial perturbation amplitude.}
\label{fig:meta}
\end{center}
\end{figure}

To show metastability, a long-lived mode, identified by the dashed curve (1a) in Fig.~\ref{fig:meta}, is perturbed at various points along its trajectory with $n=1$ perturbations of varying amplitude. The relevant transport coefficients for this baseline, metastable state were: $S=10^6,$ thermal conductivity $\kappa_\perp=4.75\times 10^{-6},$ viscosity $\mu=1.0\times 10^{-5},$ and the initial equilibrium was perturbed with the $n=1$ linear eigenfunction using a perturbation amplitude of $\epsilon = 1.0\times 10^{-6}$ (The exact meaning of this parameter is not as important as its relative amplitude.). The same equilibrium, when perturbed with $\epsilon=2\times 10^{-6}$, evolves into an explosive mode (curve 1b). Similar numerical experiments are performed at later points in the nonlinear evolution of the baseline LLM: At $t=1012,~\epsilon=4.0\times 10^{-6}$ (curve 2a) is stable, however $\epsilon=5.0\times 10^{-6}$ becomes explosively unstable (curve 2b). At a later time,  $t=1256,$ the same perturbation ($\epsilon=5\times 10^{-6}$)  leads to large-amplitude, damped oscillations but the mode remains stable (curve 3a). A slightly larger perturbation ($\epsilon=6\times 10^{-6}$) again becomes explosively unstable (curve 3b). 

Clearly, the quasi-equilibrium states representing the early nonlinear phase of the long-lived mode are metastable with respect to finite (as opposed to infinitesimal) perturbations. Transition to the explosive regime requires a perturbation amplitude above a threshold. The critical amplitude increases in time, which is expected since the background pressure profile relaxes due to dissipation, thus gradually reducing the free-energy source for the mode.

\section{One dimensional model}

In this section we present a simple system that exhibits linear instability, metastability and  explosive behavior, which should make the nonlinear results of the previous sections more intuitive and easier to understand. Of course the results of this one dimensional model are not meant to be a quantitative explanation for the complex nonlinear behavior seen with the full MHD equations.

\begin{figure}[htbp]
\begin{center}
\includegraphics[width=6in]{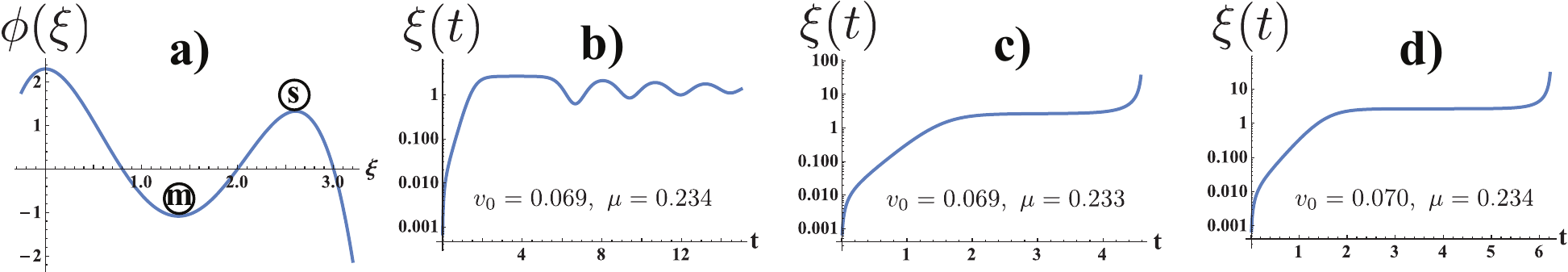}
\caption{\em \baselineskip 14pt (a) Potential $\phi(\xi)$ used in the $1D$ model. $\xi=0$ is the linearly unstable equilibrium point. The points labelled \textcircled{m} and \textcircled{s} denote the metastable equilibrium and the bifurcation saddle points, respectively. (b) For $v_0\equiv {\dot{\xi}(0)}=0.069,~\mu=0.234$ we have damped oscillations that lead to a ``long-lived'' stable state.  Slightly lower dissipation, $\mu=0.233,$ in (c) or  slightly higher initial velocity, $v_0=0.070,$ in (d) both lead to explosive behavior with a finite-time singularity.}
\label{fig:OneD}
\end{center}
\end{figure}

We start with the $1D$ equation of motion for a particle moving in a potential $\phi$ and experiencing a damping force:
\begeqn
{\ddot \xi} = -\frac{\partial \phi}{\partial \xi} - \mu{\dot\xi},~\xi(0) = 0,~{\dot \xi(0)} = v_0 \ge 0,
\endeqn
where $\mu$ is the damping coefficient. We choose the potential to be the quartic polynomial $\phi(\xi) = -(\xi-c_0)(\xi-c_1)(\xi-c_2)(\xi-c_3),$ where $c_0=-0.48,~c_1=0.8,~c_2=2,~c_3=3$. As seen in Fig.~\ref{fig:OneD} (a), the point $\xi=0$ represents a linearly unstable equilibrium (we will consider only $\xi\ge 0$ here). With any positive velocity perturbation ($v_0 > 0$) the particle will move down the potential hill, exhibiting ``linear instability.'' But if the damping coefficient is large enough ($\mu > \mu_{cr}(v_0)$), it will not be able to climb out of the well and eventually settle at the metastable equilibrium point marked with \textcircled{m} at $\xi=1.39$. This behavior is shown in Fig.~\ref{fig:OneD} (b), where, after a period of exponential growth, the  displacement $\xi(t)$ exhibits damped oscillations.
However, with only slightly lower dissipation, the particle is able to move past the saddle point \textcircled{s} at $\xi=2.60$ and become ``explosively unstable,'' as seen in panel (c). Of course the same result can be achieved by keeping the dissipation level constant but increasing the initial perturbation (panel (d)).

The damped oscillations of Fig.~\ref{fig:OneD} (b) correspond to the long-lived mode described by curve (1) in Fig.~\ref{fig:bifurcation} (a). The explosive instability in Fig.~\ref{fig:OneD} (c) that develops at a lower dissipation level corresponds to the curves (2) and (3) of Fig.~\ref{fig:bifurcation} (a). Finally the explosive instability in Fig.~\ref{fig:OneD} (d) that results from a higher initial velocity corresponds to the explosive curve $1b$ in Fig.~\ref{fig:meta} that follows a larger perturbation of the initial equilibrium. Curves $2b, 3b$ of Fig.~\ref{fig:meta} have not been simulated with the $1D$ model, but clearly they correspond to large-velocity perturbations of the damped oscillations in Fig.~\ref{fig:OneD} (b).

\section{Summary and discussion}

In this work we demonstrate that the nonlinear evolution of a pressure-driven $n=1$ kink-ballooning mode can exhibit a bifurcation between a benign final state with little confinement degradation--a long-lived mode (LLM),  and an explosive instability that results in a  fast disruption with very short precursors. The bifurcation depends sensitively on assumed transport levels and the initial perturbation amplitude. Large diffusive transport or too small a perturbation leads to a saturated $n=1$ LLM. Equivalently, there is a transport-dependent critical perturbation amplitude, $v_{cr}=v_{cr}(\eta,\mu,\kappa_\perp,\ldots)$, such that $v > v_{cr}$ leads to explosive behavior.
The long-lived mode itself is metastable and can be pushed into the explosive regime, again with a finite perturbation above a threshold. Thus it is possible that a LLM can abruptly terminate with a fast disruption.

Since a benign LLM is a possible end state, it is clear that the initial $n=1$ instability has to be weak and not too far from an instability threshold; a robust and ideally unstable $n=1$ is unlikely to saturate without serious deleterious effects on confinement. Thus we have concentrated on weak, resistive modes far from ideal instability boundaries. This choice follows also from the expectation that an MHD mode does not come into existence as a robustly unstable ideal mode with a large growth rate; the resistive thresholds tend to be much lower and the instability generally appears first as a weak, resistive mode. But this feature (weak instability) that makes a LLM possible would at the same time seem to make it difficult to explain a fast disruption with little or no precursors, the other possible end state of the bifurcation. 

This difficulty is resolved by the numerical observation, with some theoretical support (e.g., \cite{cowley2015}), that a weakly growing $n=1$ mode can become explosive nonlinearly. Thus, a feeble resistive instability can transform into a robust ideal mode in a short period of time.  Although a detailed understanding of the nonlinear process responsible for this transformation is lacking at this point,  a simple one-dimensional model is presented that mimics essentially all its important features: a linear instability that can either saturate in a metastable state or lead to a nonlinear explosive instability.  
Of course the most important consequence of this explosive instability is that it obviates the need for a long period of mode evolution on transport time scales where it gradually becomes stronger with the changing background equilibrium. Thus a long series of precursor oscillations are avoided. 
 
One particular feature of the pressure-driven $n=1$ mode that plays an important role in the transition to the explosive regime is the quadrupole geometry of the pressure perturbation due to the dominant $m/n=2/1$ harmonic. This perturbation naturally adds an elliptical deformation to the flux surfaces, which can nonlinearly turn into a ballooning finger. Once formed, the finger rapidly moves outward, pushing through flux surfaces while essentially maintaining its original $2/1$ symmetry. To minimize bending of the field lines as it moves into regions with $q\gg 2$, it becomes localized both in parallel and perpendicular directions. Thus, although it is originally quite extended along the field lines, it turns into a highly local nonlinear structure as it moves to  the edge, transporting a significant portion of the energy in the core to a small area on the wall. Although not shown here but discussed elsewhere\cite{aydemir2016}, the rapid ejection of the core is accompanied by stochastization of the field outside the finger, while the finger itself remains well-confined by regular flux surfaces. These features are consistent with jet-like flows in some high-$\beta$ disruptions in TFTR\cite{fredrickson1996} and JET\cite{paley2005}. It is likely that the global $n=1$ mode responsible for the well-documented fast high-$\beta$ disruption in DIII-D\cite{chu1996, jayakumar2002} also has similar origins.

Finally, ITER disruption mitigation efforts seem to be based on the anticipation that a resistive wall mode or a tearing mode would slowly lock to the wall prior to the disruption, giving at least a few tens of milliseconds of warning time. Without a perfect  pressure-profile control system, fast disruptions of the type discussed in this work would make the efficacy of this approach questionable in some of the advanced scenarios planned for ITER.

\section{Acknowledgements}
This work was supported by MSIP, the Korean Ministry of
Science, ICT and Future Planning, through the KSTAR project.

\end{document}